\documentclass[preprint,nofootinbib,amsmath,amssymb,aps,preprintnumbers,superscriptaddress]{revtex4-1}
\usepackage{graphicx} % Required for inserting images
\usepackage{amsmath,amsthm,latexsym,amscd,amsfonts,bm,hyperref}
\usepackage{here}
\usepackage{mathrsfs}
\usepackage{xcolor}
\begin{document}

\preprint{KOBE-COSMO-25-07}

\title{Graviton-photon conversion in blazar jets as a probe of high-frequency gravitational waves}

\author{Himeka Matsuo}
\email[]{himeka.matsuo@grad.nao.ac.jp}
\affiliation{Astronomical Science Program, Graduate Institute for Advanced Studies, SOKENDAI, 2-21-1 Osawa, Mitaka, Tokyo 181-8588, Japan}
\affiliation{Department of Physics, Kobe University, Kobe 657-8501, Japan}

\author{Asuka Ito}
\email[]{asuka@phys.sci.kobe-u.ac.jp}
\affiliation{Department of Physics, Kobe University, Kobe 657-8501, Japan}

%\date{March 2025}

\begin{abstract}
We study graviton-photon conversion in the magnetic fields of a blazar jet and explore the possibility of detecting high-frequency gravitational waves through blazar observations.
We calculate the conversion rate using the magnetic field configurations of leptonic, lepto-hadronic, and hadronic one-zone synchrotron self-Compton models for the blazar jet of Mrk 501.
By requiring that the photon flux produced within the blazar jet does not exceed the observed flux of Mrk 501, we derive conservative constraints on the abundance of stochastic gravitational waves.
We find that, for all three models considered, the resulting limits can be more stringent than previous constraints in the frequency range from $10^8$ Hz to $10^{15}$ Hz.
%In particular, in the hadronic model, we obtain 
%$h^2\Omega_{{\rm GW}} \lesssim \mathcal{O}(1)$ around GHz.
\end{abstract}

\maketitle

%\tableofcontents
\section{Introduction}

A gravitational wave from a binary black hole merger was observed by LIGO in 2015~\cite{LIGOScientific:2016aoc}. 
This was the first observation of gravitational waves, which opened the era of the gravitational wave astronomy/cosmology. 
The frequency of the gravitational waves detectable with LIGO is
around kHz.
Pulsar timing arrays reported the evidence of the signal from stochastic gravitational waves in the frequency range of around nHz~\cite{NANOGrav:2023gor,EPTA:2023fyk}.
Therefore, multi-frequency gravitational wave observations have already begun, enabling us to explore new physics through the detection of various gravitational wave sources.
In this regard, developing the methods for observing high-frequency 
gravitational waves above $10$kHz is demanded to probe interesting high-frequency gravitational wave sources~\cite{Aggarwal:2025noe,Ghiglieri:2015nfa,Wang:2019kaf,Ghiglieri:2020mhm,Ringwald:2020ist,Ghiglieri:2022rfp,Vagnozzi:2022qmc,Ema:2021fdz,Ema:2015dka,Ema:2016hlw,Ema:2020ggo,Nakayama:2018ptw,Huang:2019lgd,Barman:2023ymn,Khlebnikov:1997di,Easther:2006gt,Easther:2006vd,Garcia-Bellido:2007nns,Ito:2016aai,Aggarwal:2020olq,Franciolini:2022htd,Saito:2021sgq,Gehrman:2022imk,Gehrman:2023esa,
Casalderrey-Solana:2022rrn,Ito:2025lcg,Dong:2015yjs}.

One way to detect high-frequency gravitational waves is to use tabletop size detectors~\cite{Ito:2019wcb,Ito:2020wxi,Berlin:2021txa,Ito:2022rxn,Ito:2023bnu,Berlin:2023grv,Bringmann:2023gba,Kanno:2023whr}, 
since the sensitivity of a detector is expected to be maximized when its typical size is comparable to the wavelength of a target gravitational wave.
In particular, axion observations can be utilized for
high-frequency gravitational wave searches as well~\cite{Ito:2019wcb,Ejlli:2019bqj,Domcke:2022rgu,Domcke:2022rgu}.
Along with the rapid development of quantum sensing, pushing forward the study of tabletop size gravitational wave detectors would be an interesting and important direction~\cite{Ito:2023bnu}.

Another way to probe high-frequency gravitational waves is to utilize the electromagnetic wave probe, whose typical frequency range is $10^8$Hz$\sim10^{35}$Hz, with the use of graviton-photon conversion~\cite{Pshirkov:2009sf,Dolgov:2012be,Domcke:2020yzq,Ramazanov:2023nxz,Liu:2023mll,ito2024probing,ito2024gravitational,Dandoy:2024oqg,He:2023xoh,Lella:2024dus,McDonald:2024nxj,Kushwaha:2025mia,Li:2025eoo}.
Graviton-photon conversion inevitably occurs in the presence of a background magnetic field.
It implies that gravitational waves in the universe 
convert to electromagnetic fields universally, since
various magnetic fields exist across a wide range of scales such as intergalactic magnetic fields, galactic magnetic fields, and magnetic fields around stars.
Possibilities for probing high-frequency gravitational waves with
graviton-photon conversion in the intergalactic magnetic fields~\cite{Domcke:2020yzq,ito2024gravitational,He:2023xoh,Kushwaha:2025mia,Li:2025eoo}, 
galactic magnetic fields~\cite{ito2024gravitational,Lella:2024dus}, pulsar magnetic fields\cite{ito2024probing,Dandoy:2024oqg,McDonald:2024nxj}, and geomagnetic fields~\cite{Liu:2023mll,ito2024gravitational} have been studied so far.
In this paper, we study graviton-photon conversion
within the magnetic fields of a blazar and give 
new constraints on stochastic gravitational waves by using observational data of Mrk 501.

Mrk 501 is a blazar, a type of active galactic nucleus (AGN) that features an astrophysical jet~\cite{NASAIPAC}.
Considering gravitons which propagate along the jet, they can be converted into photons 
in the jet magnetic fields. 
We employ three jet models, leptonic, lepto-hadronic, and hadronic one-zone synchrotron self-Compton (SSC) to calculate the graviton-photon conversion rates within the jet. 
By comparing the expected electromagnetic signals produced by gravitational waves in the jet magnetic fields with observational data of telescopes, we derive new constraints on stochastic gravitational waves, which can be stringent than previous works in the frequency range of $10^8$–$10^{15}$ Hz.
Then, the possibility of using this method to probe certain gravitational wave sources is briefly discussed.

The organization of the paper is as follows.
In Sec.\ref{Grav-photon}, we briefly explain graviton-photon conversion in the presence of 
background magnetic fields and plasmas.
In Sec.\ref{sec3_1}, we introduce three kinds of one-zone SSC models, 
which effectively describe the magnetic field configuration of jets.
In Sec.\ref{sec3_2}, we obtain conservative upper limits on stochastic gravitational waves by comparing the 
calculated photon flux converted in the blazar jet of Mrk 501 with the observed photon spectra from Mrk 501.
The final section is devoted to the conclusion.

\section{Mixing between gravitons and photons}\label{Grav-photon}
In this section, we briefly review the graviton-photon conversion
in the presence of a background magnetic field~\cite{Raffelt:1987im}.
We consider the following action:
\begin{align}\label{action}
  S=\int d^4x\sqrt{-g} \left[ \frac{M^2_{pl}}{2}R-\frac{1}{4}F_{\mu\nu}F^{\mu\nu} \right],
\end{align}
where $M_{pl}$ is the reduced Planck mass, $R$ is the Ricci scalar, $g$ is the determinant of the metric $g_{\mu\nu}$. The field strength of electromagnetic fields is defined by $F_{\mu\nu}=\partial_\mu \mathscr{A}_\nu-\partial_\nu \mathscr{A}_\mu$ with the vector 
potential $\mathscr{A}_\mu$.

Let us expand the vector potential and the metric as
\begin{align}
  \mathscr{A}_{\mu}(x)&=\bar{A}_{\mu}+A_{\mu}(x),\\
  g_{\mu\nu}(x)&=\eta_{\mu\nu}+\frac{2}{M_{pl}}h_{\mu\nu}(x),
\end{align}
where $\bar{A}_\mu$ consists of background magnetic fields $\bar{B}^i = \epsilon^{ijk}\partial_j\bar{A}_k$,  
$\eta_{\mu\nu}$ is the Minkowski metric, and $h_{\mu\nu}(x)$ is a traceless-transverse tensor.
Below, we adopt the gauge $A_0=0$ and consider only the two transverse modes, as the longitudinal mode (if any) is irrelevant to our study.
Now, we consider gravitons and/or photons propagating along $z$-direction. 
We also take the direction of the background magnetic field to be $y$-direction, which is orthogonal to the propagation direction, i.e., $\boldsymbol{\bar{B}}=(0,\bar{B},0)$.
We can also take the polarization bases for the vector and the tensor as
\begin{align}
\label{bases}
  e^+_i=
  \begin{pmatrix}
  1 \\
  0 \\
  0
  \end{pmatrix},\enspace
  e^\times_i=
  \begin{pmatrix}
  0 \\
  1 \\
  0
  \end{pmatrix},\enspace
  \epsilon^+_{ij}=\frac{1}{\sqrt{2}}
  \begin{pmatrix}
  1 & 0 & 0 \\
  0 & -1 & 0 \\
  0 & 0 & 0
  \end{pmatrix},\enspace
  \epsilon^\times_{ij}=\frac{1}{\sqrt{2}}
  \begin{pmatrix}
  0 & 1 & 0 \\
  1 & 0 & 0 \\
  0 & 0 & 0
  \end{pmatrix}.
\end{align}
Here, two linear polarizations of photons are leveled by $+$ and $\times$ according to 
plus and cross modes of gravitational waves.
The electromagnetic field and the gravitational wave can be expanded as follows
\begin{align}\label{A}
  A_i&= e^{-i(\omega t-kz)}A^\sigma(z)e^\sigma_i \, , \\\label{h} 
  h_{ij}&= e^{-i(\omega t-kz)}h^\sigma(z)\epsilon^\sigma_{ij} \, ,
\end{align}
with the bases (\ref{bases}) for $\sigma=+,\times$. 
$A^\sigma(z)$ and $h^\sigma(z)$ are the amplitudes of the electromagnetic wave and the gravitational wave, respectively.
They can vary during their propagation due to the graviton-photon mixing.
Using Eqs.(\ref{action})-(\ref{h}) with the implementation of
effects of magnetized plasmas and higher order contributions in QED~\cite{ito2024probing},
one can derive
\begin{align}\label{Ahplusmode}
  \left[i\partial_z+
  \begin{pmatrix}
  -\frac{1}{2\omega}\frac{\omega^2\omega_{p,(i)}^2}{\omega^2-\omega^2_{c,(i)}}+\frac{1}{2\omega}\frac{16\alpha^2\bar{B}(z)^2\omega^2}{45m_e^4}+\frac{1}{2\omega}\frac{88\pi^2\alpha^2T^4\omega^2}{2025m^4_e} & -i\frac{\bar{B}(z)}{\sqrt{2}M_{pl}}
\\
  i\frac{\bar{B}(z)}{\sqrt{2}M_{pl}} & 0 
  \end{pmatrix}
  \right]
  \begin{pmatrix}
    A^+(z)\\
    h^+(z)
  \end{pmatrix}
  \simeq0 , 
\end{align}
\begin{align}
  \label{Ahcrossmode}
  \left[i\partial_z+
  \begin{pmatrix}
  -\frac{\omega^2_{p,(i)}}{2\omega}+\frac{1}{2\omega}\frac{28\alpha^2\bar{B}(z)^2\omega^2}{45m_e^4}+\frac{1}{2\omega}\frac{88\pi^2\alpha^2T^4\omega^2}{2025m^4_e} & -i\frac{\bar{B}(z)}{\sqrt{2}M_{pl}} \\
  i\frac{\bar{B}(z)}{\sqrt{2}M_{pl}} & 0 
  \end{pmatrix}
  \right]
  \begin{pmatrix}
    A^\times(z)\\
    h^\times(z)
  \end{pmatrix}
  \simeq0 . 
\end{align}
In deriving Eqs.~(\ref{Ahplusmode}) and (\ref{Ahcrossmode}), we assumed that the scale of graviton-photon conversion is much larger than $k^{-1}$, and that photons are ultrarelativistic, i.e., $\omega \simeq k$. We also neglected the spatial derivatives of the magnetic field $\bar{B}$.
$T$ denotes the temperature of the cosmic microwave background.
The plasma frequency $\omega_{p}$ and the cyclotron frequency $\omega_{c}$ are defined by
\begin{align}\label{pl}
  \omega_{p,(i)}&=\sqrt{\frac{4\pi\alpha n_{i}}{m_{i}\gamma_{i}}},\\
  \omega_{c,(i)}&=\frac{eB(z)}{m_i\gamma_{i}},
\end{align}
where $\alpha$ is the fine structure constant, $n_e$ ($n_p$) is the electron (proton) number density, and
$m_e$ ($m_p$) is the electron (proton) mass.  
$\gamma_e$ ($\gamma_p$) represents the Lorentz factor of electrons (protons), which
suppresses the plasma frequencies since the effective masses of the plasma particles get heavier due to
the Lorentz factors.
Although the mixing between electromagnetic waves and gravitational waves is caused by the off-diagonal terms in Eqs.~(\ref{Ahplusmode}) and (\ref{Ahcrossmode}), the conversion efficiency is also affected by the diagonal terms, which correspond to the effective masses of the photons.
Therefore, not only the magnitude and distribution of background magnetic fields, but also plasma properties such as number densities should be carefully investigated.
As long as the magnitude of the magnetic field is not so large, 
we can ignore the contribution of the cyclotron frequency in Eq.(\ref{Ahplusmode}) 
as $\omega_c/\omega \ll  1$ holds. 
Then, the plus and cross modes experience the same plasma effect approximately.
Therefore, there is no actual difference in the conversion rates of the plus and cross modes 
in the low-frequency range where the plasma effect dominates the high order QED effect.
On the other hand, in the high-frequency range where the high order QED effect is dominant, 
the conversion rates of plus and cross modes can become slightly different.
In the next section, we study graviton-photon conversion in the magnetic fields of a jet by modeling the distributions of magnetic fields and plasma particles, with a particular focus on Mrk 501.

\section{Graviton-photon conversion in a blazar jet}\label{sec3}

It is known that magnetic fields exist on various scales in our universe.
We then expect that graviton-photon conversion occurs everywhere, and such phenomena 
may be used for probing high-frequency gravitational waves~\cite{Pshirkov:2009sf,Dolgov:2012be,Domcke:2020yzq,Ramazanov:2023nxz,Liu:2023mll,ito2024probing,ito2024gravitational,Dandoy:2024oqg,He:2023xoh,Lella:2024dus,McDonald:2024nxj,Kushwaha:2025mia,Li:2025eoo}.
In the previous works, graviton-photon conversion in intergalactic regions~\cite{Domcke:2020yzq,ito2024gravitational,He:2023xoh,Kushwaha:2025mia,Li:2025eoo}, in the Milky Way Galaxy~\cite{ito2024gravitational,Lella:2024dus}, within magnetic fields around stars~\cite{Liu:2023mll,ito2024probing,ito2024gravitational,Dandoy:2024oqg,McDonald:2024nxj}
have been studied.
In this section, we investigate graviton-photon conversion in the jet magnetic field of an 
AGN with particular attention on Mrk 501, which has been continuously monitored for over a 
decade~\cite{Abe_2023}.
Then, we calculate the photon flux converted from stochastic gravitational waves in the jet of Mrk 501,
compare it with the observed data of photon flux from Mrk 501 to assess the sensitivity of this method.

\subsection{Modeling of magnetic fields and plasma}\label{sec3_1}
Mrk 501 is a blazar located about 119\,Mpc away ($z = 0.034$), 
hosting a central black hole with a mass of $\sim10^9$ solar 
mass~\cite{NASAIPAC}.
Although observations have not revealed the magnetic field configuration in the jet of Mrk 501 completely yet, 
it has been indicated that the helical component, which is almost orthogonal to jet direction, 
is dominant~\cite{Hu:2024eqt} as is common for BL Lacertae objects~\cite{Marscher:2021ntl}.
This implies that graviton-photon conversion can occur effectively during the propagation within the blazar jet.
In practice, an effective model of the magnetic field configuration is often employed to fit the spectral energy distribution (SED) of the target blazar.
The one-zone synchrotron self-Compton (SSC) model is often 
adopted for this effective modeling~\cite{Tavecchio:1998xw}.
In the one-zone SSC model, relativistic charged particles emit synchrotron radiation and upscatter these photons through inverse Compton scattering, all within a single emission region of radius $r_{{\rm em}}$, which is filled with a tangled and homogeneous magnetic field of strength $B_{{\rm em}}$, 
meaning that the magnitude and directional dependence are averaged, in the jet's comoving frame.
Outside the emission region, one would expect that magnetic fields expand freely without radiative pressure.
Thus, the strength of the magnetic field, envisioned as being dominated by its helical component, would scale as
\begin{align}
    B(r)&=B_{{\rm em}}  \hspace{2.25cm} \textrm{for}\enspace  r\leq r_{{\rm em}}, \label{B1}\\
    B(r)&= B_{{\rm em}}\left(\frac{r}{r_{{\rm em}}}\right)^{-1} \quad \textrm{for}\enspace r_{{\rm em}}<r . \label{B2}
\end{align}
Note that all quantities characterizing the jet models discussed here and in the following are measured in the comoving frame of the jet.
Since a jet exhibits relativistic bulk motion, with a Lorentz factor typically of $\Gamma\sim$ a few tens. This implies that the emission from the one-zone modeled jet is highly beamed with an opening angle of $\sim 1/\Gamma$ for an observer, although it is assumed to be isotropic in the jet’s comoving frame.
The $z$-axis used in Eqs.~(\ref{Ahplusmode}) and (\ref{Ahcrossmode}) corresponds to the line of sight and is taken to lie within the opening angle.
We assume that the above scaling of the magnetic field continues 
to $r_{{\rm end}}$ at which its amplitude becomes comparable to 
the typical galactic magnetic fields, $\sim 1\mu$G.
Below this field strength, the influence of the surrounding environment outside the jet may become significant, and the simple scaling may no longer be applicable.
%
%\begin{figure}[H]
%  \centering
%  \includegraphics[width=100.0truemm]{configuration.png}
%  \caption{three dimensional configuration}
%  \label{fig:3d-configuration}
%\end{figure}
%

In addition to the magnetic field configuration, the energy fraction of plasmas in the jet and its energy spectrum must also be determined in order to reproduce the observed SED of photons from the jet.
Here, we consider three types of models for reproducing the SED of emitted photons: 
leptonic, lepto-hadronic, and hadronic models, based on~\cite{Abe_2023}.
In the leptonic model, the emission is mainly attributed to synchrotron radiation and inverse Compton scattering by leptons only, while hadronic and lepto-hadronic models assume that relativistic protons also contribute to the emission.
Furthermore, the hadronic model assumes that the high-energy component of SED, typically extending from X-rays to gamma rays, is dominated by hadron-initiated emission processes, such as proton synchrotron radiation and photomeson-induced cascades~\cite{Cerruti:2020lfj}.
Thus, a larger magnetic field strength is needed in the hadronic model
compared to the other models.

In these three kinds of the one-zone SSC models, 
the energy distributions of electrons and protons are taken 
to be~\cite{Abe_2023}
\begin{align}
    N_e(\gamma) &= N_{0,e}\cdot\gamma^{-\alpha_{1,e}},\hspace{2cm} \textrm{for}\enspace 
                   \gamma_{{\rm min},e} \leq \gamma \leq \gamma_{{\rm br}}\, ,
                 \label{N1}\\
    N_e(\gamma) &= N_{0,e}\cdot\gamma^{-\alpha_{2,e}}\gamma_{{\rm br}}^{\alpha_{2,e}-\alpha_{1,e}},\quad \textrm{for}\enspace \gamma_{{\rm br}} < \gamma \leq \gamma_{{\rm max},e}  \, ,\label{N2}
\end{align}
and 
\begin{equation}
    N_p(\gamma) = N_{0,p}\cdot\gamma^{-\alpha_p},\quad \textrm{for}\enspace \gamma_{{\rm min},p} \leq \gamma \leq \gamma_{{\rm max},p}\, ,
                 \label{Np}
\end{equation}
respectively.
$N_{0,i}$ represents a normalization constant, 
$\alpha_{1,e}$ and $\alpha_{2,e}$ are the spectral indices below and above the break energy $\gamma_{{\rm br}}$ of the electron population.
$\alpha_{p}$ is the spectral index of the proton population.
$\gamma_{{\rm min},i}$ and $\gamma_{{\rm max},i}$ represent the minimum and the maximum Lorentz factors of electrons or protons.
Using Eqs.(\ref{N1})-(\ref{Np}), one can calculate the 
averaged Lorentz factor:
\begin{align}
    \bar{\gamma}_i = \frac{\int\gamma N_i(\gamma) d\gamma}{\int N_i(\gamma) d\gamma}.
\end{align}
Moreover, in the one-zone SSC models, it is assumed that the ratio between the energy density of electrons, 
$U_e$ (protons, $U_p$), to that of the magnetic field, $U_B$, is constant.
This allows us to describe
the distribution of the number density of electrons/protons are related to the magnetic field configuration as
\begin{align}
  n_e(r)&=\frac{B(r)^2}{2\bar{\gamma}_e m_e}\frac{U_e}{U_B} , \label{ne}\\
  n_p(r)&=\frac{B(r)^2}{2\bar{\gamma}_p m_p}\frac{U_p}{U_B} .  \label{np}
\end{align}

The modeling of the jet configuration presented so far is general and not specific to Mrk 501.
The parameter values in each model are determined to fit the SED of the blazar under study.

In the next subsection, we focus on Mrk 501 and compute the graviton–photon conversion probability using the one-zone SSC modeling of the jet configuration consistent with its SED. 
This treatment allows us to reduce the problem to one dimension, as a consequence of averaging the magnetic field distribution in the jet’s comoving frame. 
We believe that this approximation is valid for estimating the order of magnitude. 
Moreover, a more effective graviton–photon conversion would be expected if the helical component is dominant in an actual jet, which may indeed be favored~\cite{Hu:2024eqt,Marscher:2021ntl}. 
Therefore, compared to the case without angular averaging of the magnetic field distribution, the present one-zone modeling may provide a conservative estimate of the graviton–photon conversion probability for leptonic, lepto-hadronic, and hadronic models.

\subsection{Constraining gravitational waves with Mrk 501 observations}\label{sec3_2}
When gravitational waves pass through the magnetic fields in a jet,
the gravitational waves convert into electromagnetic waves.
In principle, the converted photons can be observed with telescopes which are observing the target jet.
Here, we focus on stochastic gravitational waves and calculate
the converted photon flux in the jet magnetic field of Mrk 501 by solving Eqs.(\ref{Ahplusmode}) and (\ref{Ahcrossmode}) with Eqs(\ref{B1}), (\ref{B2}), (\ref{ne}), (\ref{np}). 
Recalling that a black hole is located at the center of an AGN jet,
one might worry that stochastic gravitational waves could be blocked by the black hole.
However, the typical radius of the jet near the black hole is much larger than the Schwarzschild radius~\cite{Giroletti:2008xv}. 
Therefore, in practice, the existence of a black hole does not 
affect the calculation of the graviton-photon conversion in the 
jet magnetic field.
When we calculate the conversion of stochastic gravitational waves into
photons, the integration range is taken to be from $r=0$ to $r=r_{{\rm end}}$.
Moreover, as discussed in~\cite{ito2024probing}, the conversion rate
highly depends on the configuration of the magnetic field at the 
boundary, $r_{{\rm end}}$, if the oscillation length is short for the frequency
under study, and even can be zero easily.
Therefore, we introduce cutoff frequencies by assuming that both very low and very high frequencies with short oscillation lengths have zero conversion rate, and calculate only the frequencies whose oscillation length at the boundary exceeds $r_{{\rm end}}$.

Now let us evaluate the flux of the photons converted from stochastic gravitational waves.
There would exist stochastic gravitational waves in a jet, 
which can be characterized with the characteristic amplitude $h_c$ defined by~\cite{MAGGIORE2000283}
\begin{align}
  <h_{ab}(t)h^{ab}(t)>=2\int^{f=\infty}_{f=0}d(\textrm{log}f)h_c^2(f).
\end{align}
Due to graviton-photon conversion in a blazar jet magnetic field, 
photons are generated, 
its flux $F$ is given by~\cite{ito2024probing}
\begin{align}
  F = \frac{\pi^2 M^2_{pl} f h_c^2 r_{{\rm end}}^2}{d^2}(P^{(+)}+P^{(\times)})
\end{align}
where $d$ is the distance between a blazar and the Earth, $P^{(+)}$ ($P^{(\times)}$) represents conversion rate of the plus (cross) mode.
Here, no initial polarization of the gravitational waves has been assumed.
The photons are assumed to be generated isotropically in the comoving frame of the jet, since we adopt the one-zone SSC model, which assumes
an isotropic emission region, to calculate the conversion rate.
%Thus, the characteristic amplitude is represented as follows
%
%\begin{align}
%  h_c = \frac{d\sqrt{F}}{\pi M_{pl}\sqrt{f}r_{{\rm end}}\sqrt{P_{r_g-%r_{{\rm end}}}^{(+)}+P_{r_g-r_{{\rm end}}}^{(\times)}}}.
%\end{align}
It would be useful to use the energy density parameter of gravitational waves, which is related to $h_c$ as $\Omega_{{\rm GW}}(f)=\frac{2\pi^2}{3H^2_0}f^2h_c^2(f)$.

Requiring that produced photon flux within the blazar jet of Mrk 501 is 
smaller than the observed flux, one can give conservative
constraints on the abundance of stochastic gravitational waves.
%For this purpose, we use data of the SED of Mrk 501 given in Ref.~\cite{NASAIPAC}.
The parameters characterizing Mrk 501 such as magnetic field strength,
the number density of plasmas in the leptonic, lepto-hadronic, and hadronic one-zone SSC models are summarized in Table~\ref{tab:modelpar} based on Ref.~\cite{Abe_2023}.
The values of the magnetic fields already include the angular average of their components perpendicular to the gravitational wave propagation direction.
We mention that the parameters are derived for the low-state SED of Mrk 501 observed during 
2017-06-17 to 2019-07-23 (MJD 57921 to MJD 58687), so that
the contribution of transitional 
high states~\cite{VERITAS:2010vjk,VERITAS:2016xjz,MAGIC:2018uqd,Sahu:2019scf,MAGIC:2020sil} has been excluded~\cite{Abe_2023}.
\begin{table}[th]
\centering
\caption{Parameter values for the one-zone SSC models of Mrk 501. All parameters are defined in the comoving frame of the jet.\\}
\begin{tabular}{llll}
\hline
      & Hadronic    & Leptonic     & Lepto-hadronic    \\ \hline
$B_{{\rm em}}$[G]     & 3    & 0.025 & 0.025 \\
$r_{{\rm em}}$[cm]     & $1.14\times10^{17}$ & $1.14\times10^{17}$  & $1.14\times10^{17}$  \\
$\gamma_{{\rm min},e}$ & 400  & 1000  & 1000  \\
$\gamma_{{\rm br},e}$ & non  & $2.5\times10^5$ & $2.0\times10^5$  \\
$\gamma_{{\rm max},e}$ & $3.5\times10^4$  & $1.2\times10^6$ & $1.2\times10^6$ \\
$\bar{\gamma}_e$ & 1100 & 2600 & 2600 \\
$\gamma_{{\rm min},p}$ & 1    & non   & 1     \\
$\gamma_{{\rm max},p}$ & $1.1\times10^{10}$ & non & $2.0\times10^7$ \\
$\bar{\gamma}_p$ & 6.0 & non & 17 \\
$\alpha_{1,e}$ & 2.5 & 2.6 & 2.6 \\
$\alpha_{2,e}$ & non & 3.6 & 3.6 \\
$\alpha_{p}$ & 2.2 & non & 2.0 \\
$U_e/U_B$   & $6.1\times10^{-7}$  & 21    & 21.2  \\
$U_p/U_B  $ & 0.14 & non   &$ 2.3\times10^5  $ \\ \hline \\
\end{tabular}\\ \label{tab:modelpar}
\end{table}
For the data to be compared with the calculated photon flux from gravitational waves, 
we refer to~\cite{NASAIPAC} where observed SEDs with various telescopes, 
expected to correspond to the low state, are shown.

\begin{figure}[H]
    \centering
    \begin{minipage}{0.45\textwidth}
        \centering
        \includegraphics[width=100.0truemm]{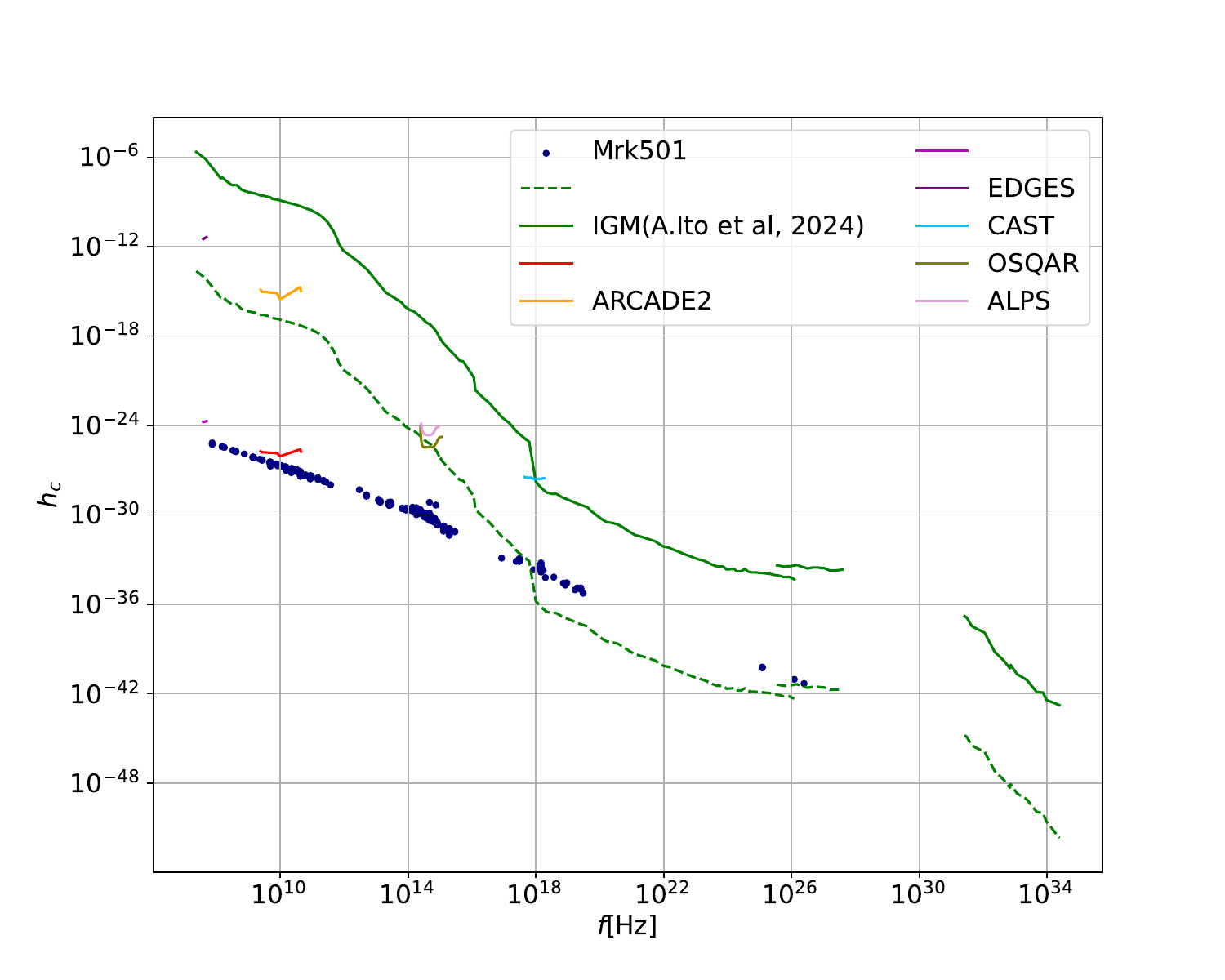}
%        \caption{(Unify three figures, write proper labels on the axes.)}
%        \label{fig:Mrk501had}
    \end{minipage}
    \hfill
    \begin{minipage}{0.45\textwidth}
        \centering
        \includegraphics[width=100.0truemm]{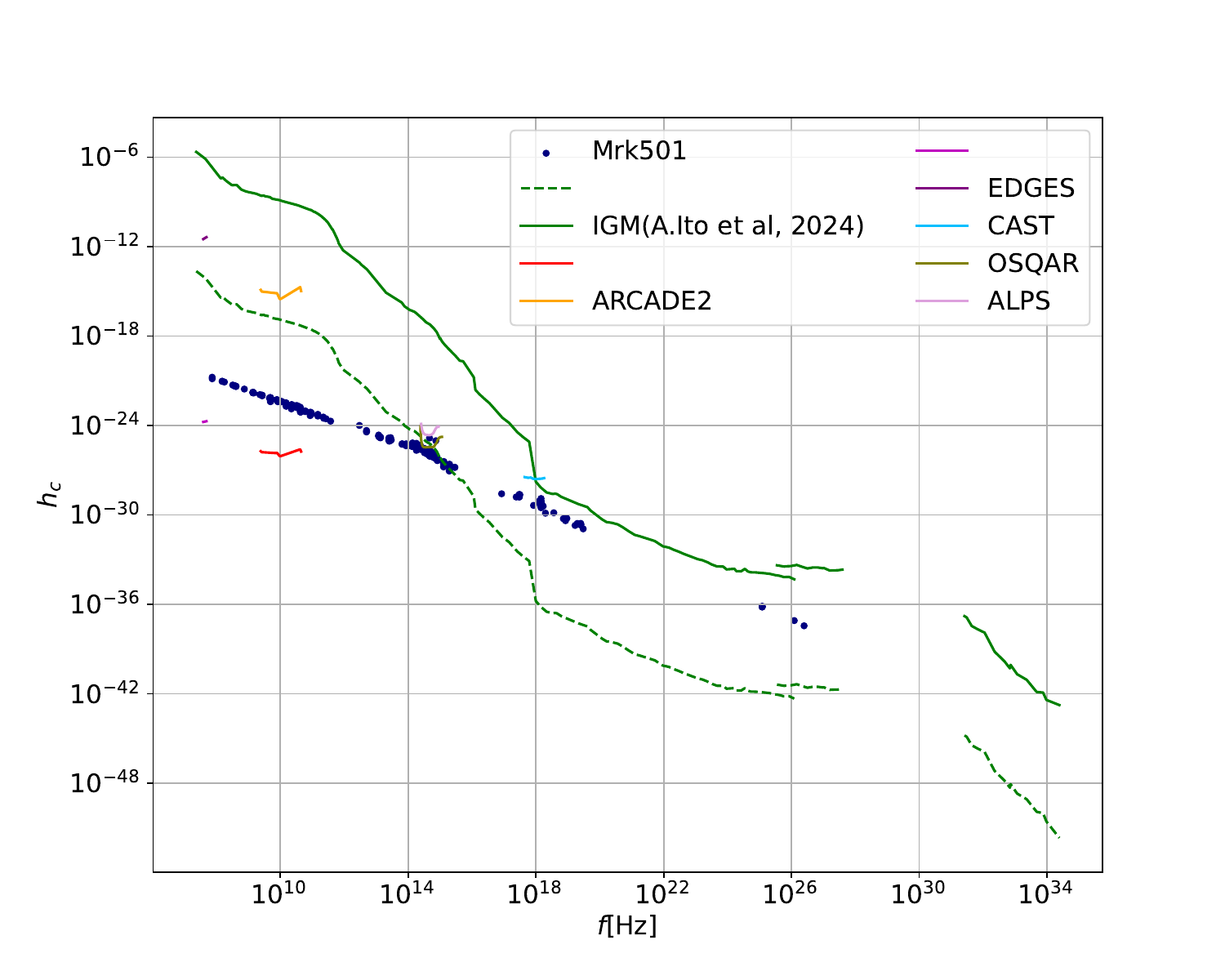}
%        \caption{(Unify three figures, write proper labels on the axes.)}
%        \label{fig:Mrk501lep}
    \end{minipage}
%\end{figure}
%
%\begin{figure}[H]
    \centering
    \begin{minipage}{0.45\textwidth}
        \centering
        \includegraphics[width=100.0truemm]{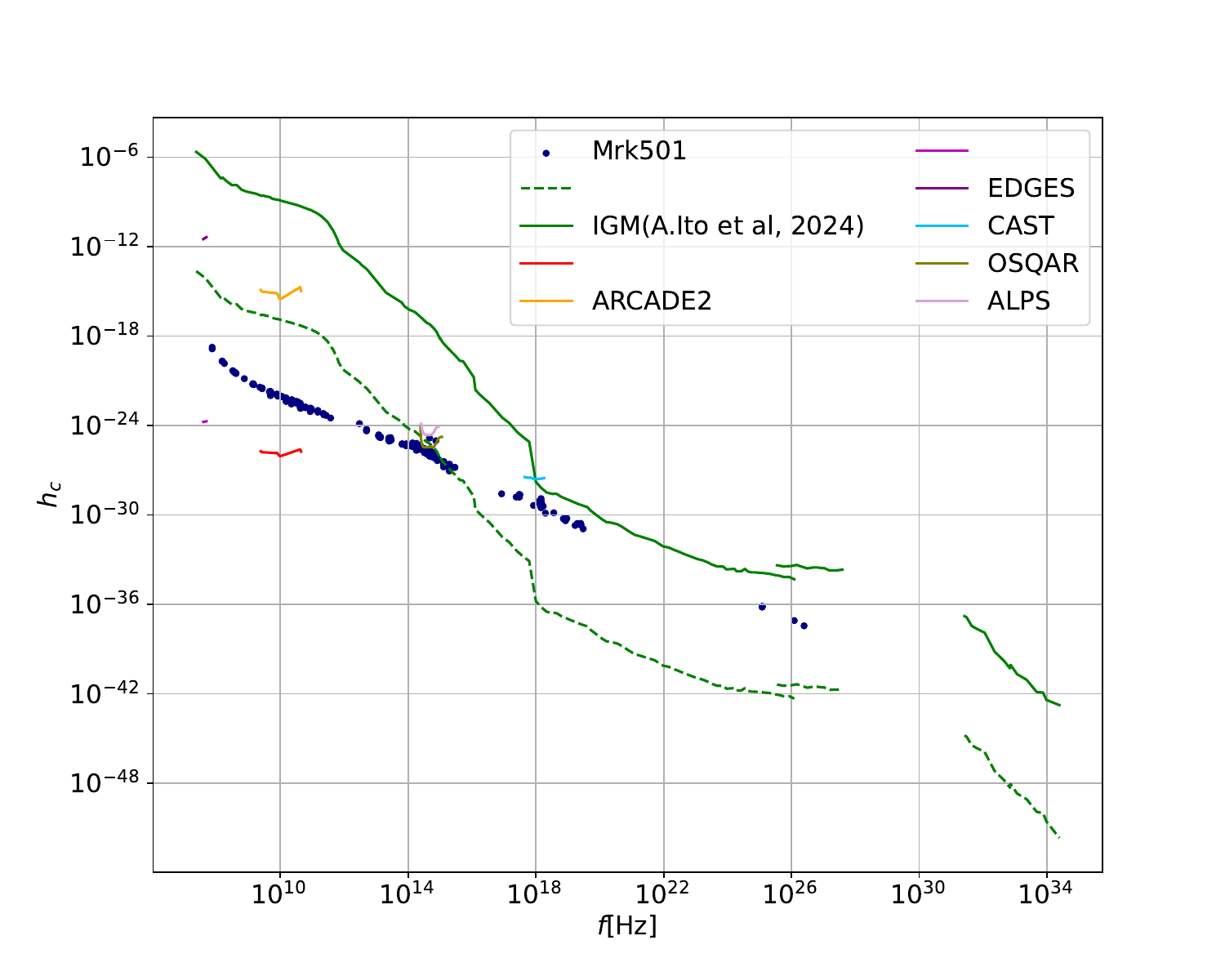}
%        \label{fig:Mrk501hc}
    \end{minipage}
    \hfill
    \begin{minipage}{0.45\textwidth}
        \centering
        \includegraphics[width=100.0truemm]{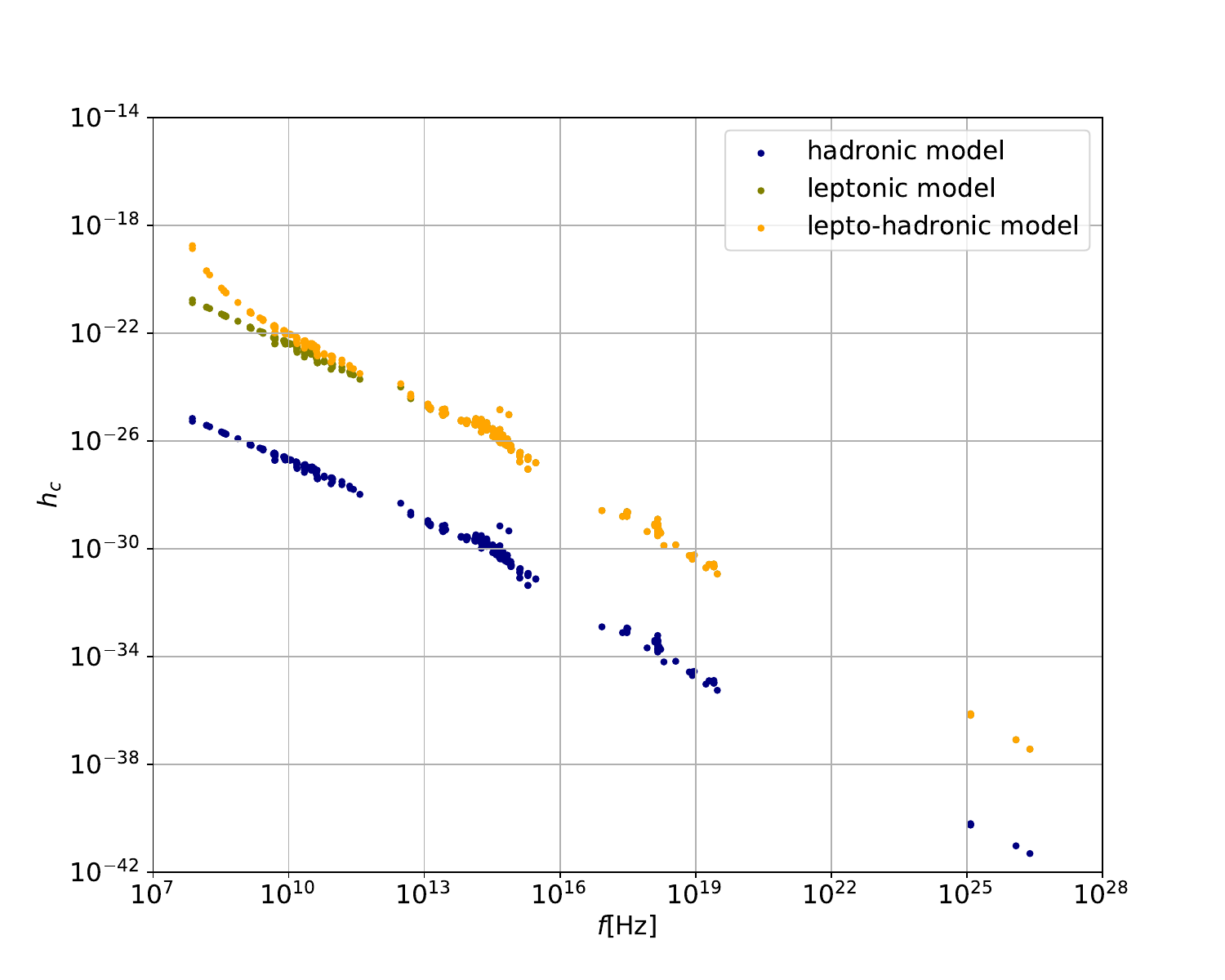}
    \end{minipage}
  \caption{Upper limits on $h_c$ of stochastic gravitational waves are shown.
The top-left, top-right, and bottom-left panels correspond to the hadronic, leptonic, and lepto-hadronic models, respectively. The bottom-right panel shows a direct comparison among the three models.
The green solid and dashed lines represent conservative and aggressive constraints, respectively, derived by considering intergalactic magnetic fields~\cite{ito2024gravitational}.
Conservative and aggressive constraints from ARCADE2 and EDGES, based on CMB spectral distortions due to graviton-photon conversion, are also shown~\cite{Domcke:2020yzq}.
Limits from axion search experiments are indicated as CAST, OSQAR, and ALPS~\cite{Ejlli:2019bqj}.
}
  \label{fig:Mrk501lh}
\end{figure}

\begin{figure}[H]
    \centering
    \begin{minipage}{0.45\textwidth}
        \centering
        \includegraphics[width=100.0truemm]{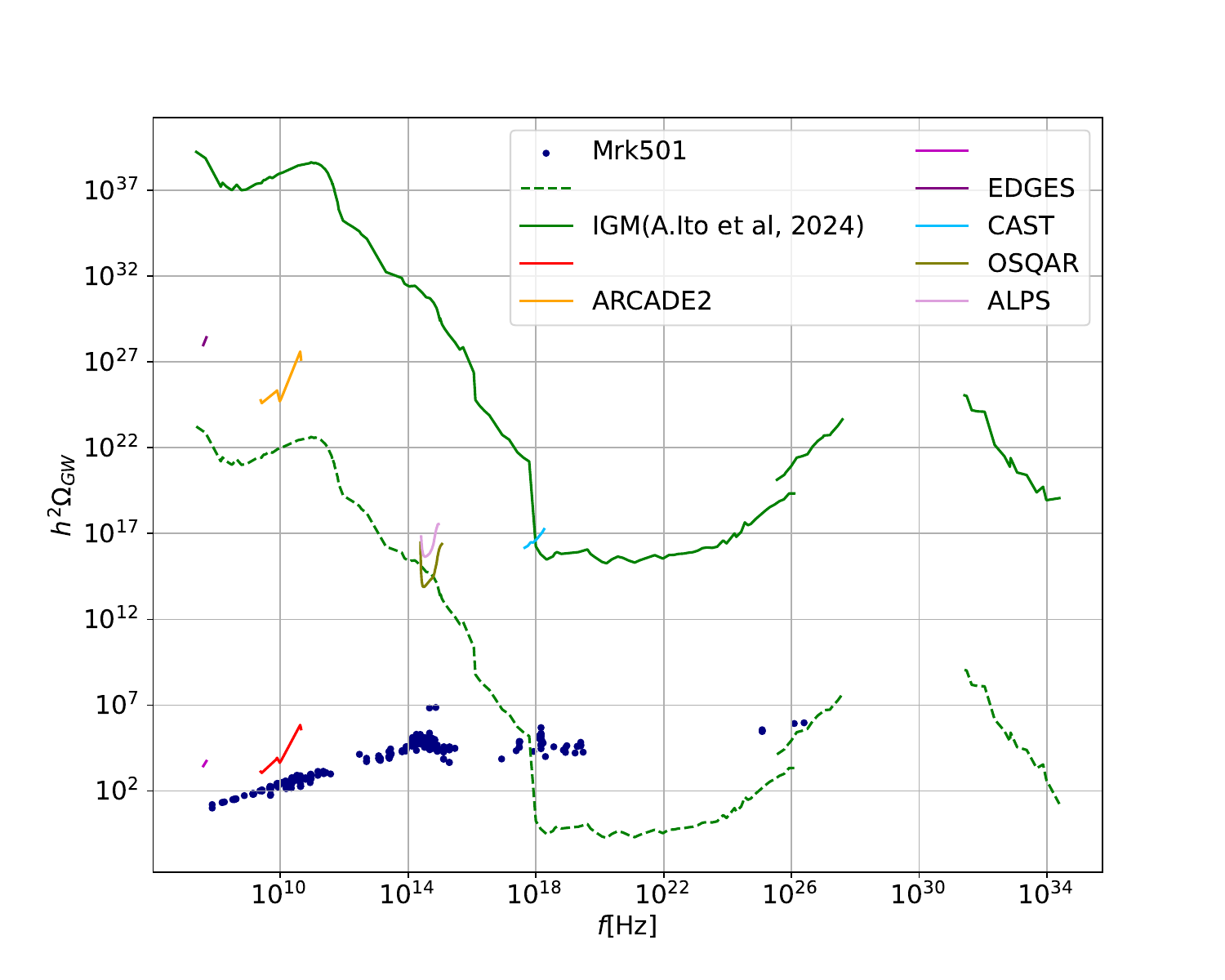}
%        \caption{(Unify three figures, write proper labels on the axes.)}
%        \label{fig:Omegahad}
    \end{minipage}
    \hfill
    \begin{minipage}{0.45\textwidth}
        \centering
        \includegraphics[width=100.0truemm]{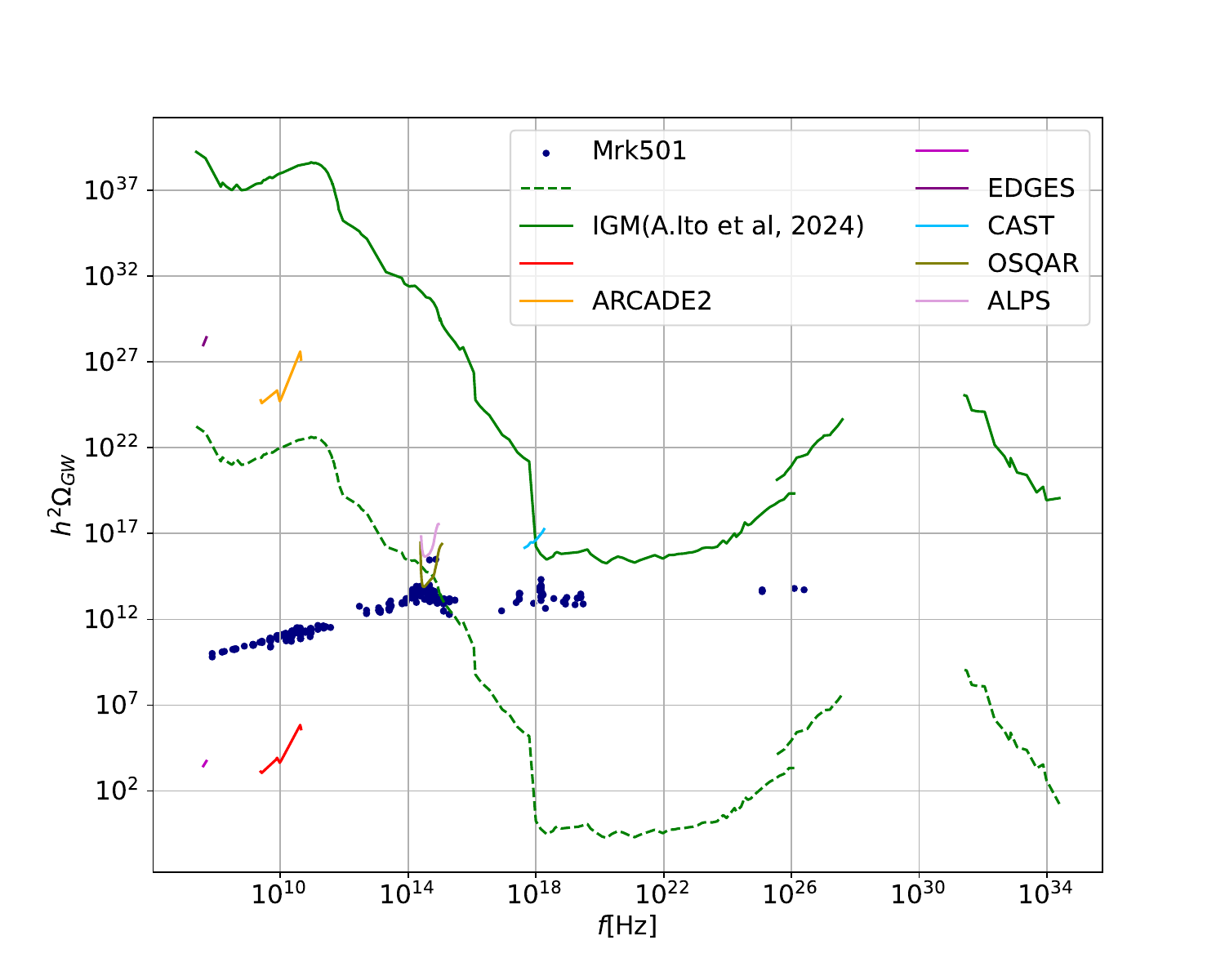}
%        \caption{(Unify three figures, write proper labels on the axes.)}
%        \label{fig:Omegalep}
    \end{minipage}
%\end{figure}
%
%\begin{figure}[H]
    \centering
    \begin{minipage}{0.45\textwidth}
        \centering
        \includegraphics[width=100.0truemm]{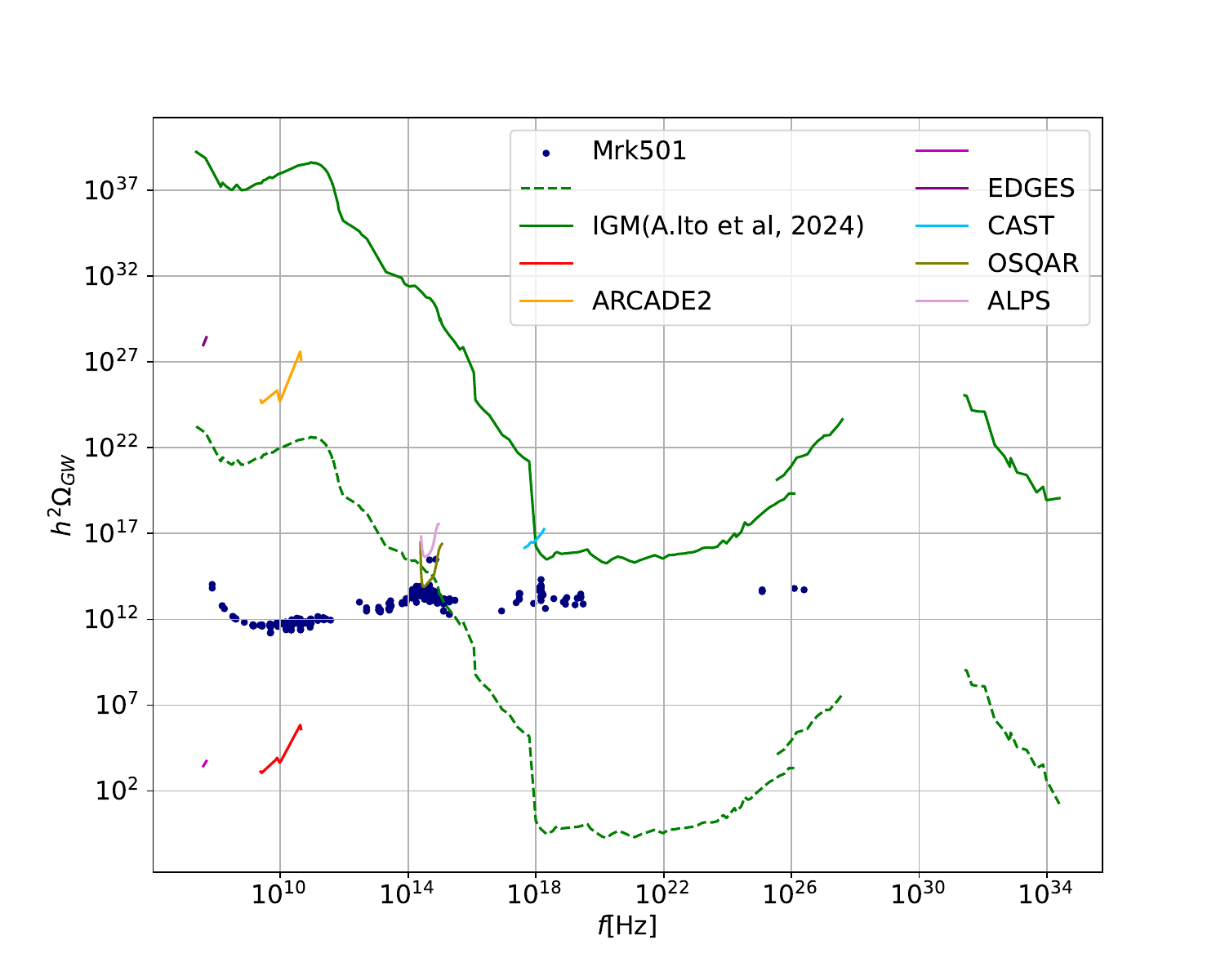}
%        \label{fig:Mrk501hc}
    \end{minipage}
    \hfill
    \begin{minipage}{0.45\textwidth}
        \centering
        \includegraphics[width=100.0truemm]{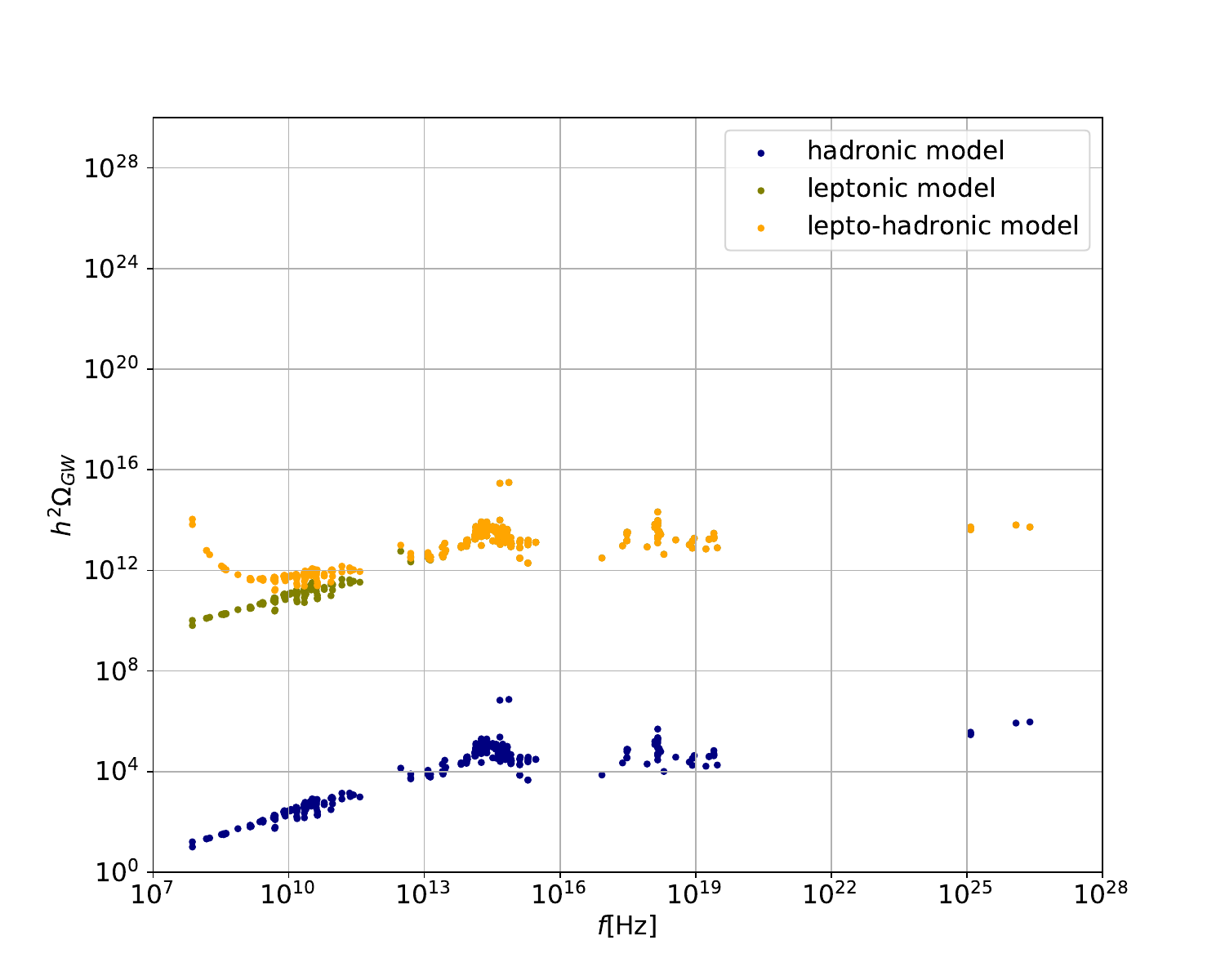}
    \end{minipage}
  \caption{Upper limits on $h^2 \Omega_{{\rm GW}}$ of stochastic gravitational waves are shown.
The top-left, top-right, and bottom-left panels correspond to the hadronic, leptonic, and lepto-hadronic models, respectively. The bottom-right panel shows a direct comparison among the three models.
The green solid and dashed lines represent conservative and aggressive constraints, respectively, derived by considering intergalactic magnetic fields~\cite{ito2024gravitational}.
Conservative and aggressive constraints from ARCADE2 and EDGES, based on CMB spectral distortions due to graviton-photon conversion, are also shown~\cite{Domcke:2020yzq}.
Limits from axion search experiments are indicated as CAST, OSQAR, and ALPS~\cite{Ejlli:2019bqj}.}
  \label{fig:Omegalh}
\end{figure}

The obtained limits on stochastic gravitational waves are shown in Figs.\ref{fig:Mrk501lh} and \ref{fig:Omegalh} for three kinds of one-zone SSC jet models.
For all three models, we obtained stringent constraints on stochastic gravitational waves in the frequency range around $10^8$-$10^{15}$~Hz, compared with, e.g., Ref.~\cite{ito2024gravitational}, where graviton-photon conversion was studied in various astrophysical magnetic fields including extragalactic regions.
Since there are uncertainties in the magnitude of extragalactic magnetic fields, the corresponding 
constraints given in~\cite{ito2024gravitational} exhibit large gaps, represented by the solid and dashed green lines in 
Figs.\ref{fig:Mrk501lh} and \ref{fig:Omegalh}.
For the same reason, constraints based on ARCADE2 and EDGES observations, which probe spectral distortions of the CMB spectrum, also have large uncertainties~\cite{Domcke:2020yzq}. 
Our results in the leptonic and lepto-hadronic models lie between the conservative and aggressive limits 
of~\cite{Domcke:2020yzq}, and provide the strongest constraints in the case of the hadronic model.
One can see that the constraints become weaker in the low-frequency tail of the lepto-hadronic model compared to the leptonic case. 
This is because the plasma effect of protons becomes significant in the low-frequency range in the lepto-hadronic case. 
However, this is not the case for the hadronic model, since the ratio of the proton energy density to the magnetic field energy density, i.e., $U_p / U_B$, is not as large in the hadronic model as in the lepto-hadronic model.
Under the hadronic model assumption, our method can probe stochastic gravitational waves with $h^2\Omega_{{\rm GW}} \sim \mathcal{O}(10)$ around the GHz range.
The reason why the conversion rate is higher for the hadronic model than for the other models is that larger magnetic fields are required to reproduce the observed SED through synchrotron radiation from protons rather than from electrons.

One potential astrophysical source that might be able to be probed using our method is the QCD bubble collisions inside neutron star merger remnants, which generate high-frequency gravitational~\cite{Casalderrey-Solana:2022rrn}.
If such a merger were to occur in Mrk~501, the resulting gravitational waves could peak at a frequency around 100\,MHz, after accounting for the Doppler shift due to a typical jet bulk Lorentz factor of a few times $10$, with a characteristic amplitude of $h_c \sim 10^{-25}$~\cite{Casalderrey-Solana:2022rrn}.
This might indicate that graviton-photon conversion in blazar jets could serve as a promising probe of such 
high-frequency gravitational wave signals originating from astrophysical phenomena in AGN jets.

\newpage
%add a 4th panel with no previous results from the literature but with all the limits

\newpage
\par

\newpage
\section{Conclusion}

We studied graviton-photon conversion in the magnetic fields of the astrophysical jet of Mrk 501. 
Using three jet models, i.e., leptonic, lepto-hadronic, and hadronic one-zone SSC models, we calculated the conversion probability within the 
jet magnetic fields.
Then, requiring that produced photon flux within the blazar jet is smaller than the observed flux of Mrk 501~\cite{NASAIPAC}, 
we gave conservative constraints on the abundance of stochastic gravitational waves.
We showed that the new limits on the stochastic gravitational waves around $10^8$-$10^{15}$ Hz can be more stringent 
than previous works.
Especially, in the hadronic model in which jet magnetic fields are stronger than those in the leptonic and lepto-hadronic models, the constraints are significantly better, $h^2\Omega_{{\rm GW}}\lesssim \mathcal{O}(10)$ around GHz.

We considered stochastic gravitational waves without specifying their origin. 
However, for a physically meaningful scenario, the gravitational waves must originate locally. 
One of the interesting candidates which may be detectable with our method is 
the high-frequency gravitational waves from QCD bubble collisions inside neutron star 
merger remnants~\cite{Casalderrey-Solana:2022rrn}.
If one considers such a neutron star merger in Mrk 501, the resulting gravitational waves could peak at a frequency around 100\,MHz, after accounting for the Doppler shift due to a typical jet bulk Lorentz factor of a few times $10$, with a characteristic amplitude of $h_c \sim 10^{-25}$~\cite{Casalderrey-Solana:2022rrn}.
On the other hand, our method in the hadronic model has a sensitivity of $h_c \sim 10^{-25}$ around 100MHz.
Although the sensitivity is to stochastic rather than transient gravitational waves, 
it might still be possible to probe such an astrophysical high-frequency gravitational wave source 
through graviton-photon conversion.

In the effective magnetic field modeling, a tangled and homogeneous magnetic field was assumed within the one-zone SSC framework. 
This model adopts a spherical emission region filled by a uniform magnetic field. 
However, in realistic situations, jets exhibit a strong collimation, 
and the corresponding magnetic fields are expected to be dominated by a helical component~\cite{Hu:2024eqt,Marscher:2021ntl}. 
%Studying graviton-photon conversion in such a setup would be interesting, 
%for example in the context of polarization effects~\cite{Kushwaha:2025mia,Li:2025eoo}, 
This may indicate that graviton-photon conversion occurs more efficiently in blazar jets, since the magnetic field is nearly perpendicular to the line of sight and tends to be amplified due to compression within the collimated jet structure.
Future evaluations of graviton-photon conversion in such a setup will be necessary for more precise estimates, 
as the structure of the jet becomes better understood.
These issues are left for future work.

\begin{acknowledgements}
A.I. was in part supported by JSPS KAKENHI Grant Number JP22K14034.
\end{acknowledgements}

\bibliography{reference.bib}

\end{document}